\documentclass[12pt,a4paper]{article}
\usepackage{graphicx}
\usepackage{amssymb,amsfonts,amsmath}
\usepackage{color}
\usepackage{cite}
\begin{document}

\newcommand{\EQ}{Eq.~}
\newcommand{\EQS}{Eqs.~}
\newcommand{\FIG}{Fig.~}
\newcommand{\FIGS}{Figs.~}
\newcommand{\TAB}{Tab.~}
\newcommand{\TABS}{Tabs.~}
\newcommand{\SEC}{Sec.~}
\newcommand{\SECS}{Secs.~}

\newcommand{\comr}[1]{\textcolor{red}{#1}}
\newcommand{\comb}[1]{\textcolor{blue}{#1}}

\title{Evolutionary dynamics and fixation probabilities in directed networks}
\author{Naoki Masuda${}^{1,2*}$ and Hisashi Ohtsuki${}^{2,3}$\\
\ \\
\ \\
${}^{1}$ 
Graduate School of Information Science and Technology,\\
The University of Tokyo,\\
7-3-1 Hongo, Bunkyo, Tokyo 113-8656, Japan
\ \\
${}^2$
PRESTO, Japan Science and Technology Agency,\\
4-1-8 Honcho, Kawaguchi, Saitama 332-0012, Japan
\ \\
${}^3$ Department of Value and Decision Science, 
Tokyo Institute of Technology,\\
2-12-1 O-okayama, Meguro, Tokyo 152-8552, Japan\\ 
$^*$ Author for correspondence (masuda@mist.i.u-tokyo.ac.jp)}

\setlength{\baselineskip}{0.77cm}
\maketitle
\newpage

\begin{abstract}
\setlength{\baselineskip}{0.77cm}
We investigate the evolutionary dynamics in directed and/or weighted networks. We study the fixation probability of a mutant in finite populations in stochastic voter-type dynamics for several update rules. The fixation probability is defined as the probability of a newly introduced mutant in a wild-type population taking over the entire population. In contrast to the case of undirected and unweighted networks, the fixation probability of a mutant in directed networks is characterized not only by the degree of the node that the mutant initially invades but by the global structure of networks. Consequently, the gross connectivity of networks such as small-world property or modularity has a major impact on the fixation probability.
\end{abstract}

\newpage

\section{Introduction}\label{sec:introduction}

Evolutionary dynamics describe the competition among different types
of individuals in ecological and social systems. Traits, either
genetic or cultural, are transmitted to others through inheritance or
imitation. The fitness of an individual determines her/his ability to
pass on her/his traits to the next generation. An individual with a
larger fitness value is more likely to replace one with a smaller
fitness value.  Such a dynamical process can be modeled using the
well-known voter model and its variants
\cite{Liggett85book,Durrettbook,Lieberman05,Antal06prl,Sood08pre}.
According to these models, individuals adopt the trait
(\textit{i.e.}, hereafter we call it type) of others.  Selection and
random drift are two major driving forces in evolutionary
dynamics. Selection results from the different fitness levels
of different
types. Random drift results from the finite size of populations.

In the voter-type dynamics, one's type is replaced with the type of
another individual.  Therefore, no new types are introduced into the
population unless an explicit mutation (or innovation) is considered.
Once a single type dominates the entire population, this unanimity
state remains the same forever.  In other words, the unanimity states
are the absorbing states of these dynamics.
Consequently, in the case of
finite populations, the stochasticity of
voter-type models leads to the fixation or extinction of
a newly introduced type after some time.
The probability that a single mutant
introduced in a population of wild-type individuals
eventually takes over the entire
population is called fixation probability \cite{Lieberman05,Antal06prl,Sood08pre,Moran58,Ewensbook,Nowak06book}.
Fixation probability quantifies
the likelihood of the propagation of a single mutant in the population.
When different types of individuals 
have the same fitness value, the resulting
evolutionary dynamics are called neutral evolutionary dynamics.
In this case, it is well-known that the fixation
probability of a single mutant on the complete graph is the reciprocal
of the population size \cite{Nowak06book}.

In reality, individuals do not necessarily interact with everyone in
the population.  They have relationships with some individuals, but
not with others. This fact leads to the idea of complex contact
networks of individuals. Neutral evolutionary
dynamics such as voter-type models in complex networks have been
extensively studied (e.g.,
\cite{Castellano03epl,Sood05prl,Suchecki05pre,Vazquez08njp}). It has
been shown that the fixation probability of a single mutant in
undirected and unweighted networks depends on the degree of the
initially invaded node as well as update rules
\cite{Lieberman05,Antal06prl,Sood08pre}.  Edges in many real networks,
however, have directionality.  Examples of real networks include
social networks 
in which a directed edge is drawn from the actor to the recipient of grooming behavior of rhesus monkeys \cite{Sade72}. Other examples
include email social networks
\cite{Ebel02_email,Newman02pre_email}, and ecological networks, in
which the heterogeneity of parameters such as habitat size
\cite{Gustafson96} and geographical biases such as wind direction
\cite{Schooley03} and reverine streams \cite{Schick07} are exemplary
sources of directionality. Moreover, the edges in real networks are
generally weighted \cite{Barrat04pnas}.
This concept has been nicely introduced in a seminal paper on 
evolutionary dynamics on graphs \cite{Lieberman05}.

In this study, we investigate the dependence of the fixation
probability of a single neutral mutant in general directed (or
weighted) networks on its initial location and on update rules. We
study three major update rules that were introduced in
\cite{Antal06prl,Sood08pre}.  Our results are remarkably different
from those obtained in the case of undirected networks in which the
fixation probability of a mutant is determined by the degree of the
node where the mutant is initially placed.  In the case of directed
networks, the fixation probability crucially depends on global
structure of networks.  
The difference between directed and undirected networks is
striking, especially in the case of 
modular, spatial, {\it or} degree-correlated
networks.

\section{Model}

Consider a population of $N$
%
%
individuals comprising two types of individuals --- type $A$ and type $B$.
Let the fitness of type $A$ 
and type $B$ individuals be r and 1, respectively.
In this study, we mainly focus on
the case $r=1$, which
corresponds to neutral competition between $A$ and $B$.
The structure of the population is described by a
directed graph $G=\{V,E\}$, where $V=\{v_1,\ldots,v_N\}$ is a set of
nodes, and 
$E$ is a set of edges, \textit{i.e.},
$v_i$ sends a directed edge
to $v_j$ if and only if $(v_i,v_j)\in E$.  Each node
is occupied by an
individual of either type. The fitness of the individual at node $v_i$
is denoted by $f_i\in \{r,1 \}$.  Each directed edge $(v_i,v_j)\in E$ is
endowed with its weight $w_{ij}$, which represents the
likelihood with which the type of individual 
at $v_i$ is transferred to $v_j$ in an update step.
We set $w_{ij}=0$ when
$(i,j)\notin E$. 

\begin{figure}
\begin{center}
\includegraphics[height=4cm,width=12cm]{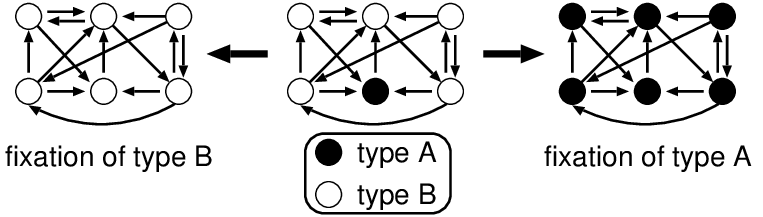}
\caption{Fixation of type $A$ or type $B$ after introduction 
of a type $A$ mutant.}
\label{fig:fixation-schem}
\end{center}
\end{figure}

Consider the introduction of a single mutant of type $A$ at node
$v_i$ in a population of $N-1$ residents of type $B$. Then, type $A$
either eventually fixates, \textit{i.e.}, takes over the entire
population, or becomes extinct (\textit{i.e.}, fixation of $B$), as
schematically shown in \FIG\ref{fig:fixation-schem}.  We are concerned
with the fixation probability of type $A$, which is denoted by $F_i$.
When needed, we refer to any one of the three update rules introduced
below by the superscript of $F_i$, such as $F_i^{LD}$ and
$F_i^{IP}$. Throughout the paper, we assume that $G$ is strongly
connected.  A network is strongly connected if there is at least one
directed path between any ordered pair of nodes.  If $G$ is not
strongly connected, we can find two nodes $v_i$ and $v_j$ such
that there is
no direct path from $v_i$ to $v_j$.  In this case, the fixation
probability $F_i$ is always zero because the individual at
$v_j$ is never replaced by the mutant initially located at
$v_i$ \cite{Lieberman05}.
Therefore, it is sufficient to investigate fixation probabilities in
the most upstream
strongly connected component of $G$. Therefore, we assume
without loss of generality that $G$ is strongly connected.

\section{Results}

\begin{figure}
\begin{center}
\includegraphics[height=12cm,width=6cm]{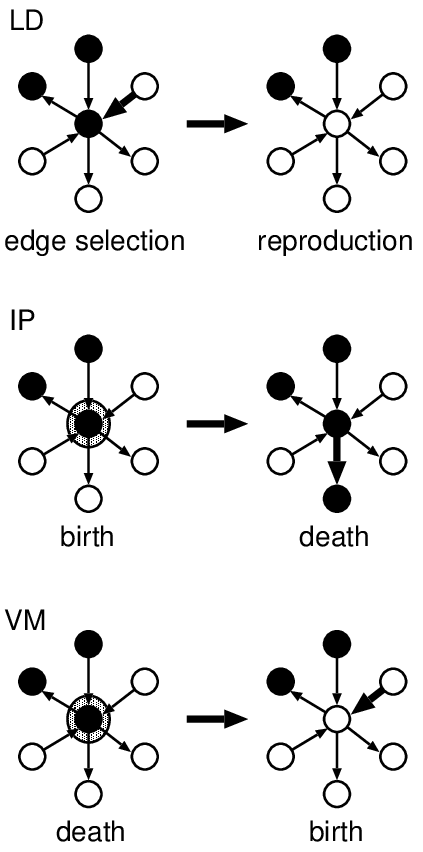}
\caption{Schematics of three update rules.}
\label{fig:ld-bd-db-schem}
\end{center}
\end{figure}

In this section, we analytically obtain a system of linear equations
that gives the fixation probabilities of mutants at individual nodes
for the three update rules (link dynamics (LD), invasion process
(IP), and voter model (VM)) \cite{Antal06prl,Sood08pre}. An update
event in the three rules is schematically
shown in \FIG\ref{fig:ld-bd-db-schem}.
Then, we compare the analytical results
with numerical results obtained for various directed networks.

\subsection{Link dynamics (LD)}

Firstly, we consider the LD
\cite{Antal06prl,Sood08pre}. In this case, one directed edge is
selected for
reproduction in each time step; the edge $(v_i,v_j)\in E$
is chosen with probability $f_i w_{ij}/\sum_{k,l}f_k w_{kl}$ for the
type of the individual at $v_{i}$ to replace the type of the individual
at $v_{j}$.
Individuals with larger fitness values and larger outgoing edge weights are
more likely to reproduce than those with smaller fitness values and
outgoing edge weights. Thus, the selection process acts on birth
events. 
Alternatively,
the selection process can be assumed to act on
death events, and the edge $(v_i,v_j)$ is chosen
with probability $(w_{ij}/f_j)/(\sum_{k,l}w_{kl}/f_l)$ for the type 
at $v_{i}$ to replace that at $v_{j}$. This
implies that
individuals with smaller fitness values 
and larger incoming edge weights are
more likely to die than those with larger fitness values and smaller 
incoming
edge weights.
In fact, the fixation probability $F_i^{LD}$ ($1\le i\le N$) is the
same under these two interpretations.

We consider the case $r=1$ (hence, $f_i=1$, $1\le i\le N$) analytically.
Suppose that a single mutant of type $A$ invades $v_i$. 
The fixation probability is given by $F_i^{LD}$.
By considering the next update event,
we can recalculate $F_i^{LD}$ as follows.
With probability $w_{ij}/\sum_{k,l} w_{kl}$,
the edge $(v_i,v_j)\in E$ is selected for reproduction.
Then, type $A$ individuals occupy $v_i$ and $v_j$. 
We denote by $F_{\{i,j\}}^{LD}$ the fixation
probability when type $A$ individuals are initially
located at $v_i$ and $v_j$
but nowhere else.
With probability $w_{ji}/\sum_{k,l} w_{kl}$,
the edge $(v_j,v_i)\in E$ is selected. Then,
type $A$ becomes extinct, and type $A$ will not
fixate. With the remaining probability
$\sum_{k\neq i,l\neq i} w_{kl}/\sum_{k,l} w_{kl}$,
the configuration of type $A$ and
type $B$ individuals does not change.  Therefore, we
obtain
\begin{equation}
F_i^{LD}=\sum_j \frac{w_{ij}}{\sum_{k,l} w_{kl}}
F_{\{i,j\}}^{LD} +
\frac{\sum_j w_{ji}}{\sum_{k,l} w_{kl}}
\times 0 +
\frac{\sum_{k\neq i,l\neq i} w_{kl}}{\sum_{k,l} w_{kl}}
F_i^{LD}.
\label{eq:Fi LD elementary}
\end{equation}
%
To prove $F_{\{i,j\}}^{LD}=F_i^{LD}+F_j^{LD}$,
consider for now
$N$ neutral types labeled 1, 2, $\ldots$, $N$ that are initially placed
at $v_1$, $v_2$, $\ldots$ $v_N$, respectively. 
On a finite graph $G$,
one of the $N$ types fixates eventually. The probability that
type $i$ or $j$ fixates is given in two ways: $F_{\{i,j\}}^{LD}$ and
$F_i^{LD}+F_j^{LD}$. This ends the proof. 
Using $F_{\{i,j\}}^{LD}=F_i^{LD}+F_j^{LD}$,
we rearrange \EQ~\eqref{eq:Fi LD elementary} as
\begin{equation}
\sum_jw_{ij}F_j^{LD} =F_i^{LD}\sum_jw_{ji}.
\label{eq:dual LD}
\end{equation}
This is a system of linear equations giving $F_{i}^{LD}$.
We note that $F_{i}^{LD}$ can be interpreted as the reproductive value of the individual at node $v_{i}$ \cite{Taylor90,Taylor96}.

Equation~\eqref{eq:dual LD} can be derived more rigorously via the
dual process \cite{Durrettbook,Donnelly83,Ewensbook,Liggett85book}.
Intuitively speaking,
the dual process of a stochastic process is another stochastic process
in which the time of the original process is reversed. The
direction of edges in the dual process is the opposite to that in the
original process.
By considering the dual process, we can understand the tree of family
lines in the original process, which is called genealogy. When we go
backward in time, two individuals sometimes `collide' in the dual
process. Such an event is called coalescence. In terms of the original
process, a coalescence corresponds to two individuals sharing the
common ancestor. After two individuals coalesce in the dual process,
they behave as a single individual, representing a single family
line.
As far as the fixation
probability is concerned, 
LD with $r=1$ is equivalent to the
continuous-time stochastic process in which each edge $(v_i,v_j)\in E$ is
selected for reproduction at the Poisson rate $w_{ij}$. 
Then, the dual process of LD
is the continuous-time coalescing random walk on
the network with all edges reversed, with a random walker
initially located on every node \cite{Durrettbook,Liggett85book}.
Coalescing random walk is defined as follows.
Consider a walker at $v_i$ moving to $v_j$ at the Poisson rate $w_{ji}$.
%
If there
is another walker at $v_j$, the two walkers coalesce into one at
$v_j$ and thereafter behave as a single random walker.
On a finite graph $G$,
the $N$ walkers eventually coalesce into one,
which is consistent with
the fact that the ancestors of all individuals
are the same in the end.
Then, $F_i^{LD}$
is the stationary density of the single random walker at $v_i$,
which is given by \EQ~\eqref{eq:dual LD}. 
As $G$ is strongly connected, the random walk on $G$ with all edges
reversed defines an irreducible Markov chain.  
Because $F_{i}^{LD}$ is the stationary density of this irreducible Markov
chain, \EQ\eqref{eq:dual LD} with constraints
$\sum_{i=1}^{N} F_{i}^{LD}=1$ and $F_{i}^{LD} \geq 0$ always has a
unique strictly positive solution.

%
%
%

The calculation of $F_i^{LD}$ from \EQ\eqref{eq:dual LD} by using
a standard
method such as the Gaussian elimination requires $O(N^3)$ computation
time. However, because relevant large
networks are usually sparse, carrying out 
the Jacobi iteration may take much less time. The convergence of
this iteration to
${\mathbf F}^{LD}$ is guaranteed by the Perron-Frobenius theorem \cite{Hornbook}.

%


In undirected graphs, $w_{ij}=w_{ji}$ holds. Therefore, $F_{i}^{LD} = 1/N$ 
solves \EQ\eqref{eq:dual LD}, giving a result previously reported in
\cite{Antal06prl,Sood08pre}. In the case of
weighted
or directed networks, the complexity of \EQ\eqref{eq:dual LD} implies that $F_i^{LD}$ is not always determined by the
local characteristics of node $v_i$
but is affected by the global structure of the networks.

Next, we argue that the mean-field (MF) approximation 
are not useful in most cases.
Consider unweighted, but possibly
directed, networks such that
$w_{ij}=1$ if $(v_i,v_j)\in E$ and $w_{ij}=0$ otherwise.
Let $k_i^{in}$ ($k_i^{out}$) be the indegree (outdegree) of $v_i$,
and we set 
\begin{equation}
\bar{F}^{LD}=\frac{1}{N}\sum^N_{i=1}F_i^{LD}.
\label{eq:barF^LD}
\end{equation}
The relation $\sum_{j} w_{ji} = k_{i}^{in}$, combined with the MF
approximation
$\sum_{i} w_{ij} F_{j}^{LD} \approx \sum_{i} w_{ij} \bar{F}^{LD} = k_{i}^{out} \bar{F}^{LD}$,
yields 
\begin{equation}
F_i^{LD} \propto \frac{k_{i}^{out}}{k_{i}^{in}}.
\label{eq:LD MF}
\end{equation}
Equation~\eqref{eq:LD MF} indicates that a large $k_i^{out}$
aids in the dissemination of the type at $v_i$ and a small
$k_i^{in}$ inhibits the replacement of the type at $v_i$ by
the type at other nodes.

\begin{figure}
\begin{center}
\includegraphics[height=6cm,width=6cm]{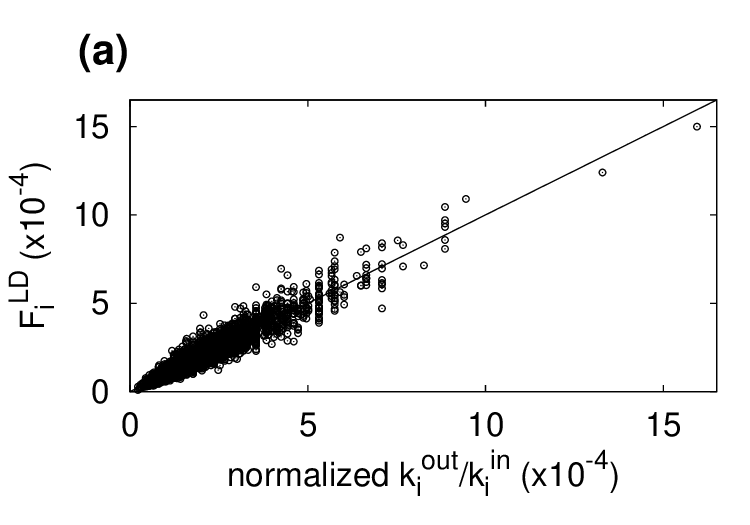}
\includegraphics[height=6cm,width=6cm]{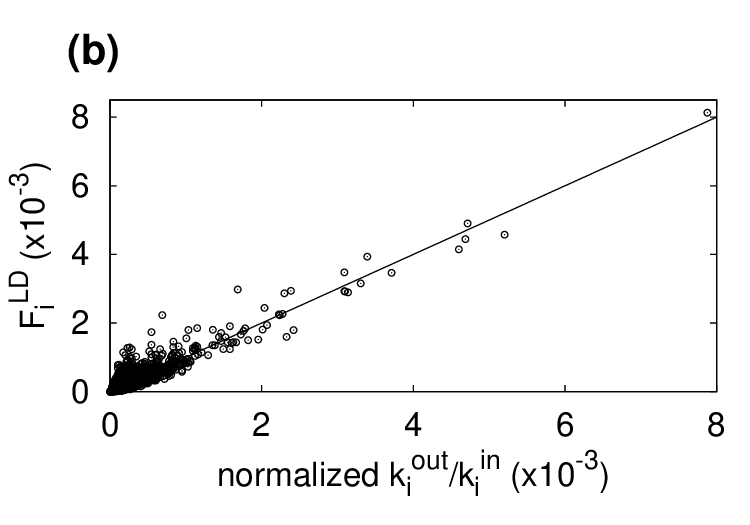}
\includegraphics[height=6cm,width=6cm]{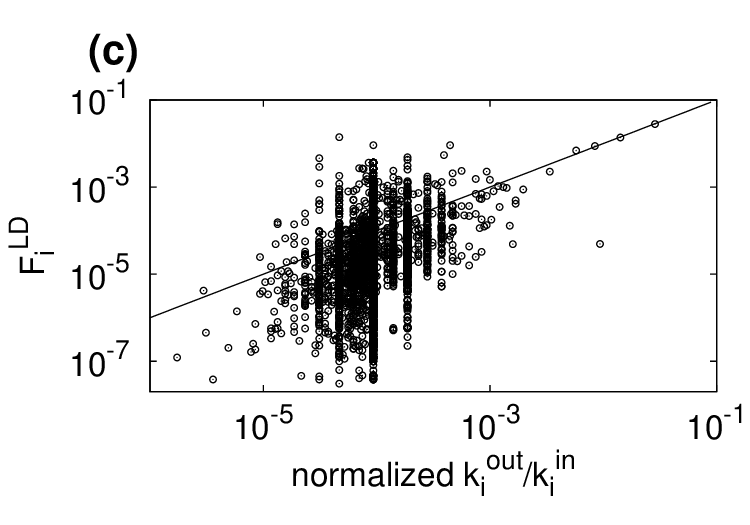}
\includegraphics[height=6cm,width=6cm]{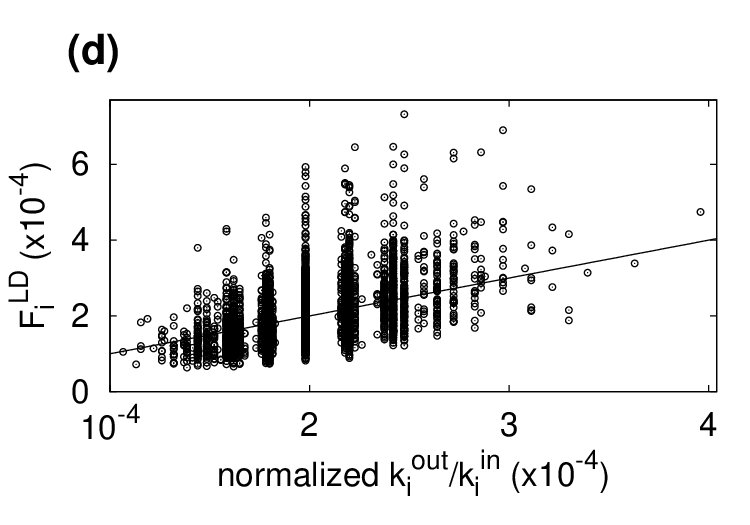}
\caption{$F_i^{LD}$ (\textit{i.e.}, the fixation probability
under LD) for a single mutant initially at node $v_i$ in
(a) an asymmetric random network with 
$N=5000$,
 (b) an asymmetric scale-free network with 
$N=5000$, (c) largest strongly connected component
of email social network with $N=9079$, and
(d) an asymmetric small-world network with $N=5000$.
The normalized $k_i^{out}/k_i^{in}$ is equal to
$(k_i^{out}/k_i^{in}) / \sum_{j=1}^N(k_j^{out}/k_j^{in})$.
The lines represent the MF ansatz
$F_i^{LD}=(k_i^{out}/k_i^{in}) / \sum_{j=1}^N(k_j^{out}/k_j^{in})$.}
\label{fig:LD}
\end{center}
\end{figure}

However, the MF approximation deviates from the correct $F_i^{LD}$ in
many cases.  As an example, consider the largest strongly connected
component of a directed and unweighted
email social network \cite{Ebel02_email} with $N=9079$
nodes and $\left<k\right>=
\left<k^{in}\right>=\left<k^{out}\right>=2.62$, where $\langle \cdot
\rangle$ denotes the average over the nodes.  In this case, $F_i^{LD}$
(indicated by the circles in \FIG\ref{fig:LD}(c)) does not agree with
the normalized $k_i^{out}/k_i^{in}$ (indicated by the line).  This is
mainly because the indegree and outdegree of the same node in this
network are positively correlated and because this network presumably
has a nontrivial global structure. Actually, the Pearson
correlation coefficient (PCC) for the $N$ pairs ($k_i^{in}$,
$k_i^{out}$), $1\le i\le N$, defined by
\begin{equation}
\frac{\frac{1}{N}\sum_{i=1}^N\left(k_i^{in}k_i^{out}-\left<k\right>^2
\right)}
{\sqrt{\frac{1}{N}\sum_{i=1}^N\left(k_i^{in}-\left<k\right>\right)^2}
\sqrt{\frac{1}{N}\sum_{i=1}^N\left(k_i^{out}-\left<k\right>\right)^2}}
\label{eq:PCC k_i}
\end{equation}
is equal to 0.40. Networks in which degrees of adjacent
nodes are correlated also show considerable discrepancies between
the MF approximation and the numerical results.
On the other hand, in the case of
undirected networks, our result $F_i^{LD}=1/N$
holds true in the presence of
degree correlation of any kind, which is consistent with
previously obtained numerical results \cite{Sood08pre}.

Next, we examine the fixation probability in
an asymmetric small-world network constructed from a
ring of $N=5000$ nodes. This network is a directed version of
the Watts-Strogatz small-world network \cite{Watts98}.
Each node of this network tentatively sends directed edges to 5 nearest
nodes along both sides. Then, 2500 out of $10N=50000$
directed edges are rewired so that
their two ends are randomly and independently
selected from the $N$ nodes, excluding 
self-loops and preexisting edges.
The correlation between the in- and out-degrees of the same node
is negligible, with the PCC for the pairs ($k_i^{in}$, $k_i^{out}$),
defined by \EQ\eqref{eq:PCC k_i},
being equal to $-0.021$. The degrees of adjacent
nodes $v_i$ and $v_j$ 
conditioned by the existence of the edge ($v_i$, $v_j$)
\cite{Newman02prl_assort,Newman03pre_mixing}
are also uncorrelated by the definition of the model.
Nevertheless, the MF approximation
is not effective in predicting the actual $F_i^{LD}$, 
as shown in \FIG\ref{fig:LD}(d).  This
discrepancy persists even for large $N$, because the
directed small-world network 
does not render $F_{i}^{LD}$ of adjacent nodes independent 
of each other. In contrast,
$F_i^{LD}$ in undirected small-world networks is completely
determined by the MF ansatz indicated by
the line in \FIG\ref{fig:LD}(d).

In contrast to these networks,
Figure \ref{fig:LD}(a) shows that the MF relation $F_i^{LD}\propto
k_i^{out}/k_i^{in}$ roughly holds for a directed random graph with
$N=5000$.  We generate a directed random graph by connecting each
ordered pair of nodes $(v_j,v_j)$ with probability
$2\left<k\right>/(N-1)$, so that
$\left<k^{in}\right>=\left<k^{out}\right>=\left<k\right>$.  In this
network, degree correlation and macroscopic network structure are both
absent, which enables the application of the MF approximation.
Even in this network, however, the MF approximation is not
exact because, as \EQ\eqref{eq:dual LD} predicts, $F_i^{LD}$ of nearby
nodes are positively correlated. Following 
\cite{Newman03pre_mixing}, we measure 
the correlation, or the assortativity, of $F_i^{LD}$ 
by the PCC for the pairs
($F_i^{LD}$, $F_j^{LD}$) for $(v_i,v_j)\in E$ defined by
\begin{equation}
\frac{\frac{1}{N}\sum_{(i,j)\in E}
\left(F_i^{LD}F_j^{LD}-\left(\bar{F}^{LD}\right)^2\right)}
{\frac{1}{N}\sum_{(i,j)\in E}^N\left(F_i^{LD}-\bar{F}^{LD}\right)^2},
\label{eq:PCC F_i}
\end{equation}
where $\bar{F}^{LD}$ is defined by \EQ\eqref{eq:barF^LD}.
The value of the PCC turns out to be slightly but
significantly positive (mean $\pm$ standard deviation based on 100
network realizations is equal to 0.0834 $\pm$ 0.0057).
%
%
The discrepancy is also significant for a large $N$, unless the mean degree
$\left<k\right>$ is large. 

The results obtained for random networks extend to the case of scale-free
networks without degree correlation.  We generate a directed scale-free
network by setting the degree distribution to be $p(k)\propto
k^{-3}$ for $k\ge \left<k\right>/2$ and
$p(k)=0$ for $k< \left<k\right>/2$, thereby generating $k_i^{in}$ and
$k_i^{out}$ ($1\le i\le N$) independently according to $p(k)$, and
randomly connecting the nodes using the Molloy-Reed algorithm
\cite{Molloy98}.  Figure~\ref{fig:LD}(b) indicates that the MF
approximation roughly explains the numerically obtained fixation
probability.

It is noted that the PCC for the pairs ($F_i^{LD}$, $F_j^{LD}$) for
$(v_i,v_j)\in E$ is small for the asymmetric scale-free network
($=0.0395\pm 0.0056$), is large for the asymmetric small-world network
($=0.7888\pm 0.0165$), and is small for the email social network
($=0.0420$).

\subsection{Invasion process (IP)}

Next, consider the IP \cite{Antal06prl,Sood08pre}.
In the IP, selection 
acts on birth.
In each time step,
$v_i$ is first selected for reproduction with probability $f_i/\sum_k
f_k$, where $f_i\in\{r,1\}$ is the fitness of the type at
node $v_i$.
Then, with probability 
$w_{ij}/\sum_l w_{il}$, the type at $v_i$
replaces that at $v_j$. Consequently, 
the probability that the edge $(v_{i},v_{j})\in E$ is used for
reproduction in an update step is equal to
$f_iw_{ij}/(\sum_k f_k \sum_l w_{il})$.
On the complete graph,
%
%
IP is the same as the standard Moran process \cite{Moran58}.
For an arbitrary $r$
the IP is mapped to the LD with the rescaled 
edge weight $w_{ij}^{\prime}=w_{ij}/\sum_{l}
w_{il}$. Therefore, from \EQ\eqref{eq:dual LD},
the fixation probability for $r=1$
is the solution to
\begin{equation}
\sum_j \frac{w_{ij}}{\sum_{l}
w_{il}} F_j^{IP} =F_i^{IP} \sum_j \frac{w_{ji}}{\sum_{l}
w_{jl}}.
\label{eq:dual IP}
\end{equation}
%
%
%
%
%
%

\begin{figure}
\begin{center}
\includegraphics[height=6cm,width=6cm]{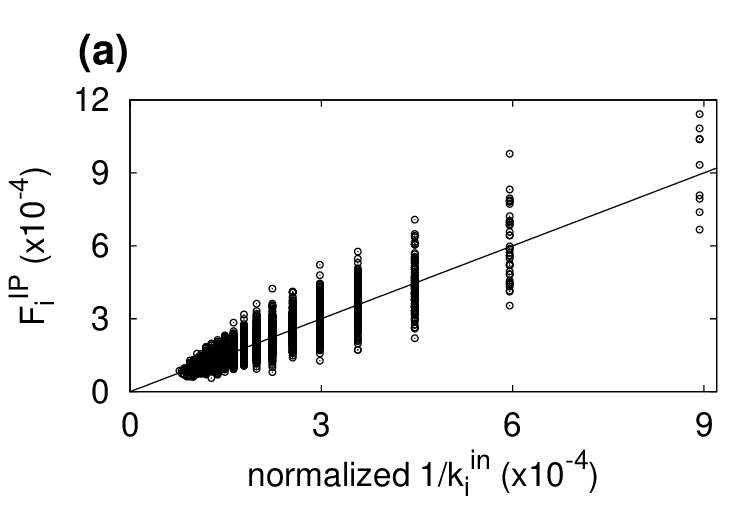}
\includegraphics[height=6cm,width=6cm]{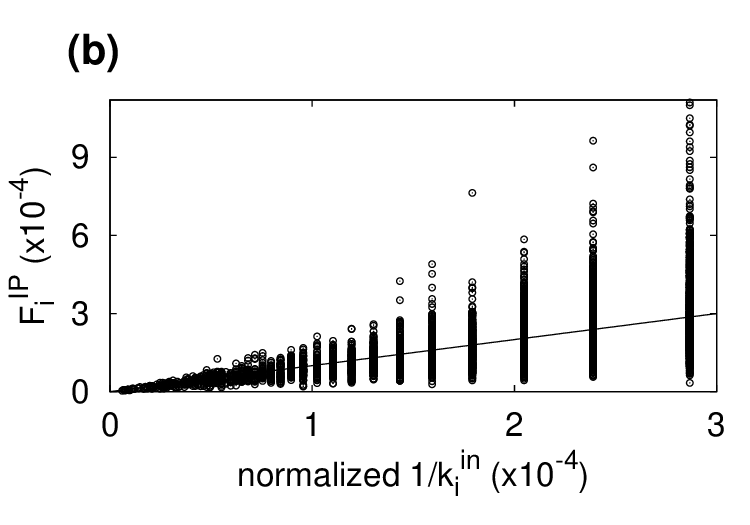}
\includegraphics[height=6cm,width=6cm]{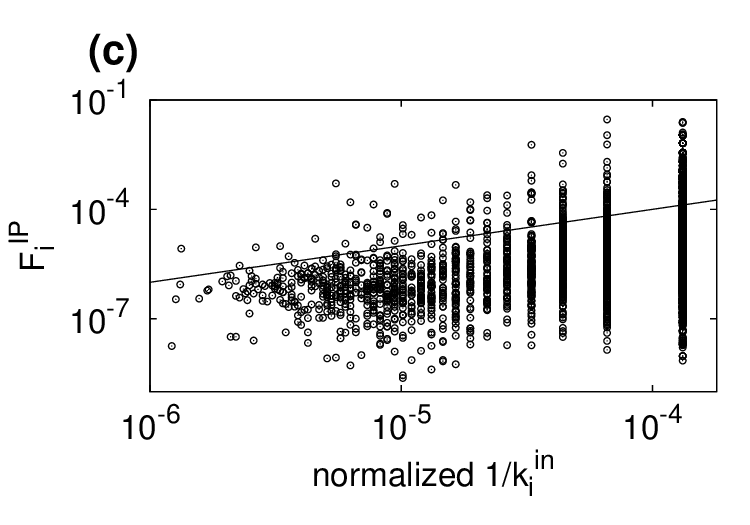}
\includegraphics[height=6cm,width=6cm]{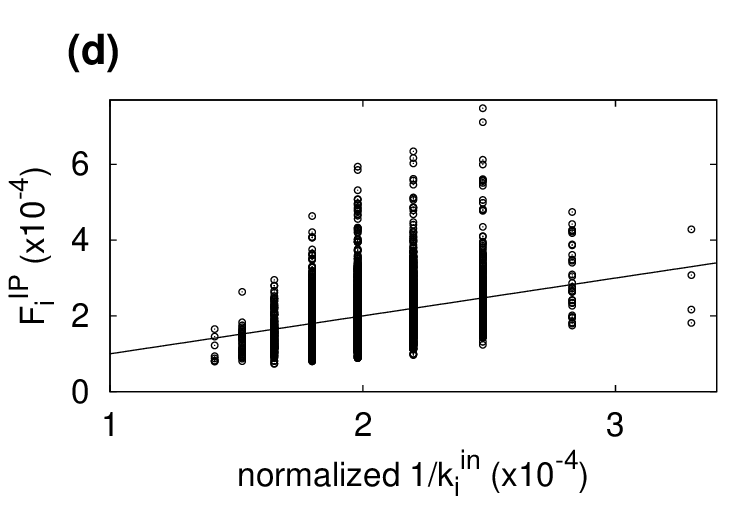}
\caption{$F_i^{IP}$ in (a) asymmetric random network,
(b) asymmetric scale-free network,
(c) email social network, and (d)
asymmetric small-world network.
The normalized $1/k_i^{in}$ is equal to
$(1/k_i^{in}) / \sum_{j=1}^N(1/k_j^{in})$.
The lines represent the MF ansatz
$F_i^{IP}=(1/k_i^{in}) / \sum_{j=1}^N(1/k_j^{in})$.}
\label{fig:IP}
\end{center}
\end{figure}

In the case of undirected unweighted networks,
$F_i^{IP}\propto 1/k_i$ solves
\EQ\eqref{eq:dual IP}, giving a previously obtained result
\cite{Antal06prl,Sood08pre,Donnelly83}. In the case of
directed unweighted
networks, applying the MF approximation to \EQ\eqref{eq:dual IP}
yields
\begin{equation}
F_i^{IP} = \frac{\sum_{j,(i,j)\in E}F_j^{IP}/k_i^{out}}
{\sum_{j,(j,i)\in E}1/k_j^{out}}
\approx \frac{(const)}{k_i^{in}}.
\label{eq:IP MF}
\end{equation}
The numerical results for the asymmetric random graph that is
used for obtaining the results shown in
\FIG\ref{fig:LD}(a) are shown in \FIG\ref{fig:IP}(a).  The
relation $F_i^{IP} \propto 1/k_i^{in}$ (the line
in \FIG\ref{fig:IP}(a)) is roughly
satisfied.  Similar to the case of LD, some deviation persists in the case of random networks even with a large $N$.  In contrast, \FIG\ref{fig:IP}(b)
indicates that the actual $F_i^{IP}$ in the scale-free network
deviates considerably from \EQ\eqref{eq:IP MF}, mainly because of the
discreteness of $k_i^{in}$ for a small integer $k_i^{in}$.
Similar to the case of LD,
the discrepancy between \EQ\eqref{eq:IP MF} and the exact
$F_i^{IP}$ is also large in the case of the email social network 
(\FIG\ref{fig:IP}(c)) and the asymmetric
small-world network (\FIG\ref{fig:IP}(d)).
In addition to the degree correlation or global structure of networks,
the discreteness of $1/k_i^{in}$ for a small $k_i^{in}$
causes further deviation, as shown
in \FIGS\ref{fig:IP}(c) and \ref{fig:IP}(d).

\subsection{Voter model (VM)}

We examine a third update rule, the so-called
voter model (VM) \cite{Antal06prl,Sood08pre}.
In the VM,
we first eliminate the type at one node $v_j$
with probability $f_j^{-1}/\sum_k f_k^{-1}$. 
Then, with probability $w_{ij}/\sum_l w_{lj}$,
the type at $v_i$ replaces that at $v_j$.
The probability that the
edge $(v_i,v_j)\in E$ is used for reproduction in an update
step is equal to
$f_j^{-1}w_{ij}/(\sum_k f_k^{-1}\sum_l w_{lj})$.
For general $r$, the VM is mapped to the LD with 
the rescaled edge weight
$w_{ij}^{\prime}=w_{ij}/\sum_l w_{lj}$. 
Consequently, from \EQ\eqref{eq:dual LD}, 
$F_i^{VM}$ for $r=1$ is given by

\begin{equation}
\sum_j \frac{w_{ij}}{\sum_{l}
w_{lj}} F_j^{VM} =F_i^{VM} \sum_j \frac{w_{ji}}{\sum_{l}
w_{li}} \;\; (= F_{i}^{VM}).
\label{eq:dual VM}
\end{equation}
%
%
%
%

\begin{figure}
\begin{center}
\includegraphics[height=6cm,width=6cm]{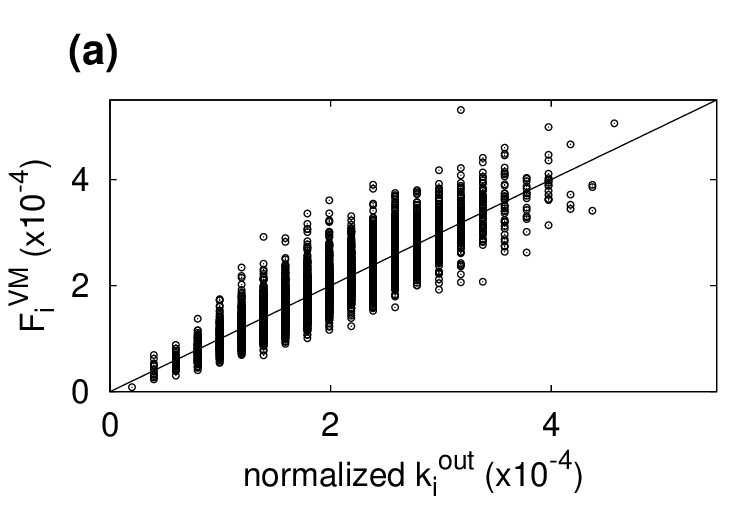}
\includegraphics[height=6cm,width=6cm]{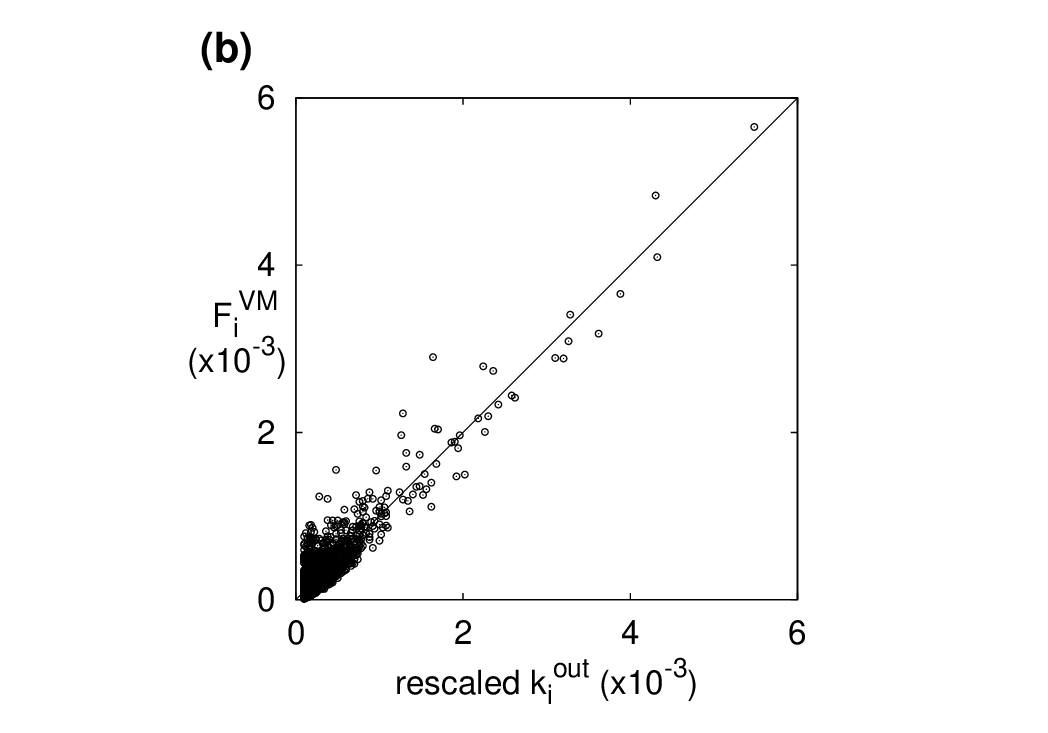}
\includegraphics[height=6cm,width=6cm]{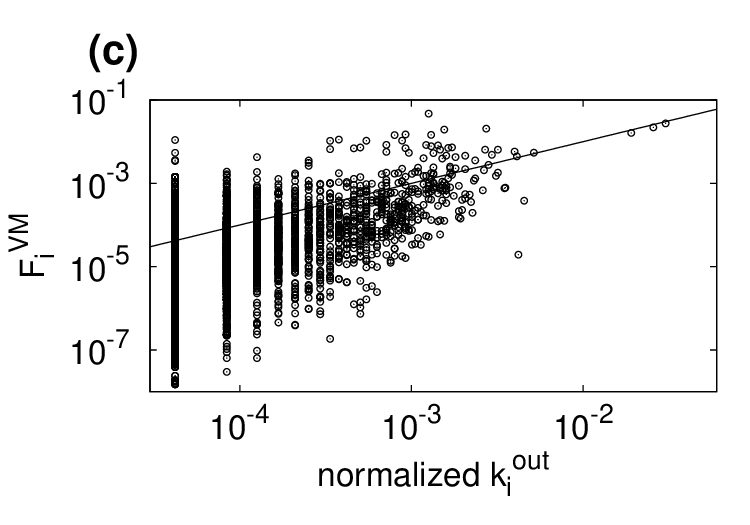}
\includegraphics[height=6cm,width=6cm]{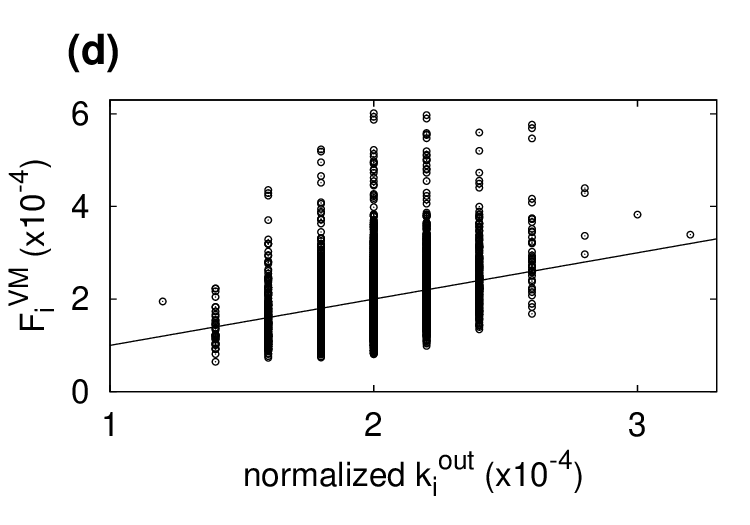}
\caption{$F_i^{VM}$ in (a) asymmetric random network,
(b) asymmetric scale-free network,
(c) email social network, and (d)
asymmetric small-world network.
The normalized $k_i^{out}$ is equal to
$k_i^{out} / \sum_{j=1}^N k_j^{out}$.
The lines represent the MF ansatz
$F_i^{VM}=k_i^{out} / \sum_{j=1}^Nk_j^{out}$.}
\label{fig:VM}
\end{center}
\end{figure}

In the case of undirected networks, $F_i^{VM}\propto k_i$ solves 
\EQ\eqref{eq:dual VM}, recovering a previously obtained result
\cite{Antal06prl,Sood08pre,Donnelly83}.
The MF approximation yields
\begin{equation}
F_i^{VM}=\frac{\sum_{j,(i,j)\in E}F_j^{VM}}{k^{in}_j}
\approx (const)\times 
k_i^{out}.
\label{eq:VM MF}
\end{equation}
In the case of the random network and the
uncorrelated scale-free network,
the numerical results shown in \FIG\ref{fig:VM}(a) and \FIG\ref{fig:VM}(b),
respectively, support the rough validity of \EQ\eqref{eq:VM MF}.
However, this naive ansatz deviates from
the actual $F_i^{VM}$ for 
the email social network (\FIG\ref{fig:VM}(c)) and
the asymmetric small-world network (\FIG\ref{fig:VM}(d)). This
situation is similar to that observed in the case of LD.

The fixation probability $F_i^{VM}$ on a graph $G$ is equivalent to
the PageRank of node $v_i$ of the graph $G^{\prime}$, where
$G^{\prime}$ is constructed by reversing all edges of $G$.  The
PageRank measures the number of directed edges a node, such as a
webpage, receives from other important nodes as exclusively as
possible \cite{Brin98,Berkhin05,Langville05}. 
If we neglect some minor technical treatments
that are necessary for the practical implementation, the PageRank
$F_i^{PR}$ of $v_i$
%
%
is defined by
\begin{equation}
\sum_j  \frac{w_{ji}}{\sum_l w_{jl}} F_j^{PR} = \lambda F_i^{PR},
\label{eq:page}
\end{equation}
where $\lambda$ is the largest eigenvalue of
the eigenequation~\eqref{eq:page}.  If many edges are directed to $v_i$,
there are many positive terms 
(\textit{i.e.}, $w_{ji}>0$) on the LHS of \EQ\eqref{eq:page}, and
they contribute to the PageRank of $v_i$ on the RHS. If $v_i$ receives
an edge from $v_j$ whose outdegree is small
or whose PageRank is large, $w_{ji}/\sum_l w_{jl}$ or $F_j^{PR}$ is
large. Each of these factors also increases $F_i^{PR}$.
A strongly connected network
yields $\lambda=1$, so that
$F_i^{PR}$ is the stationary density of the discrete-time
simple random walk on the original graph $G$
\cite{Hornbook,Brin98,Berkhin05,Langville05}.
Equation~\eqref{eq:page} with $\lambda=1$ and
with $w_{ij}$ replaced by $w_{ji}$
is identical to
\EQ\eqref{eq:dual VM}.
In PageRank, nodes that receive many edges
tend to be important, whereas the opposite is true
in the case of the VM. 
$F_i^{PR}$ is locally approximated by using
$k_i^{in}$ \cite{Fortunato06www}, which corresponds to 
the MF relation $F_i^{VM}\propto k_i^{out}$. 
However, $F_i^{PR}$
in real web graphs often deviates
from the relation $F_i^{PR}\propto k_i^{in}$ 
\cite{Donato04}. This
implies that 
$F_i^{VM}$ in real networks can also deviate from the MF
approximation, which is consistent with our main claim.

\subsection{Constant selection ($r\neq 1$)}

\begin{figure}
\begin{center}
\includegraphics[height=6cm,width=6cm]{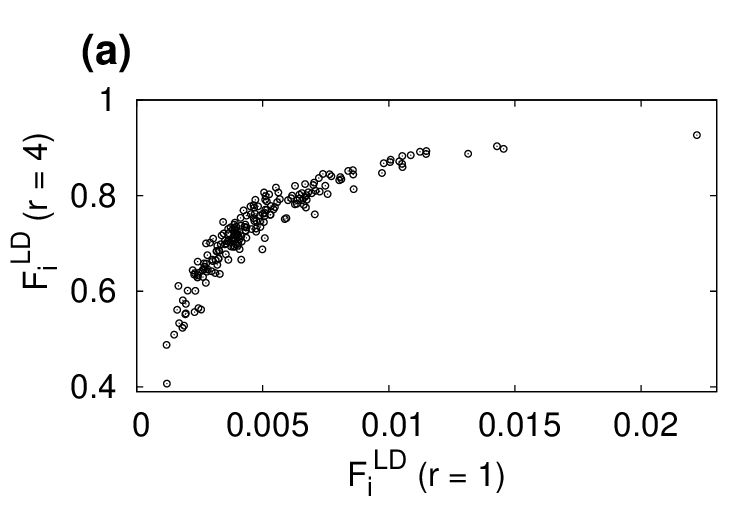}
\includegraphics[height=6cm,width=6cm]{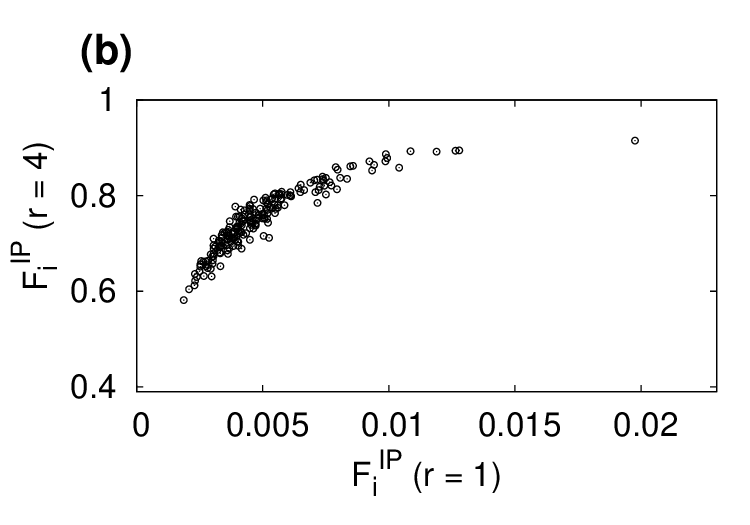}
\includegraphics[height=6cm,width=6cm]{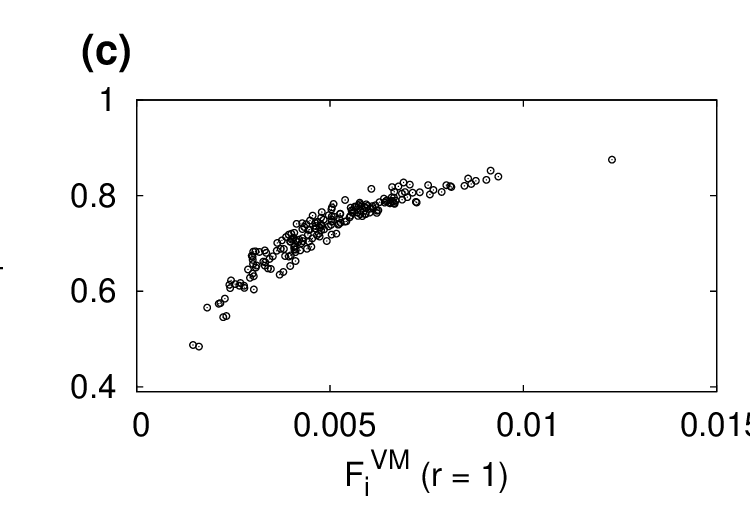}
\caption{(a) $F_i^{LD}$ for $r=4$ plotted against $F_i^{LD}$ 
for $r=1$. (b) $F_i^{IP}$ for $r=4$ plotted against $F_i^{IP}$
for $r=1$. (c) $F_i^{VM}$ for $r=4$ plotted against $F_i^{VM}$
for $r=1$. We have used an asymmetric random network with
$N=200$.} 
\label{fig:r=4}
\end{center}
\end{figure}

When $r\neq 1$, the fitness value of type A and that of 
type B are different, so
one type has a unilateral advantage over the other type. We call this
situation `constant selection'. In this case, the dual process of the
evolutionary dynamics is the coalescing and branching random
walk \cite{Durrettbook},
which is difficult to handle analytically. Therefore, we
carry out
%
Monte Carlo simulations for $r=4$ 
on a fixed
asymmetric random graph with $N=200$
and $\left< k\right>=10$.
We calculate $F_i^{LD}$ as a
fraction of runs from $2\times 10^6$ runs in which the single mutant with fitness $r$ initially located at $v_i$
(\textit{i.e.}, $f_i=r$ and $f_j=1$, $j\neq i$) 
eventually occupies all nodes of the network.
In \FIG\ref{fig:r=4}(a), the numerically obtained $F_i^{LD}$ for $r=4$ is
plotted against the exact solution of $F_i^{LD}$
for $r=1$ (\EQ~\eqref{eq:dual LD}).
Roughly speaking,
$F_i^{LD}$ for $r=4$ monotonically increases with
the exactly obtained $F_i^{LD}$ for $r=1$.
We obtain similar results in the case of the IP 
(\FIG\ref{fig:r=4}(b))
and VM (\FIG\ref{fig:r=4}(c)). 
%
Therefore, the node from which a mutant is more likely to propagate
throughout the population under the neutral dynamics ($r=1$) also
serves as a better invading node for mutants under the constant
selection ($r \in 1$). Thus, our results derived in the case of
the neutral selection is useful in predicting the order of the maginitude
of fixation probabilities under the constant selection.

\subsection{Modular networks}

Real networks are often more complex than degree-uncorrelated random,
scale-free, or small-world networks. In particular, many networks are
modular, \textit{i.e.}, they consist of several
densely connected subgraphs termed modules,
and each subgraph is connected to each other by a relatively few edges
\cite{Girvan02}. This is also the case for directed
\cite{Palla07newjp,Leicht08prl} and weighted \cite{Farkas07newjp}
networks.

\begin{figure}
\begin{center}
\includegraphics[height=6cm,width=9cm]{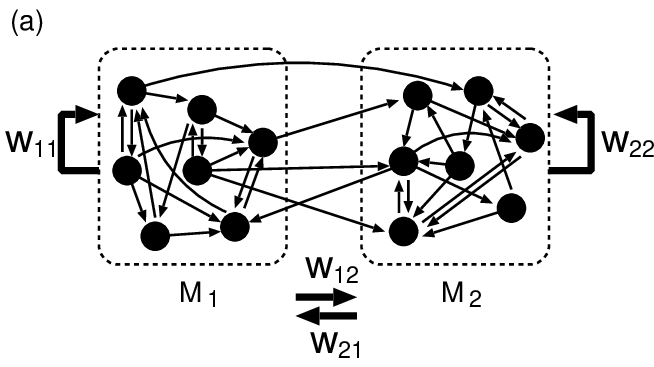}
\includegraphics[height=6cm,width=6cm]{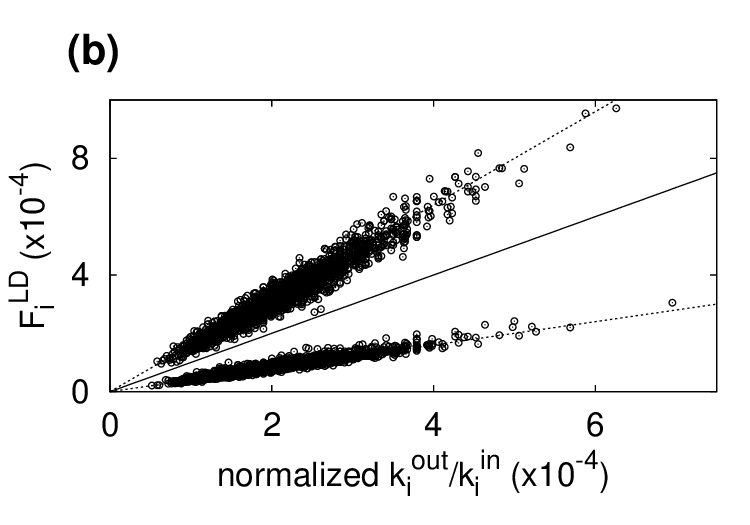}
\includegraphics[height=6cm,width=6cm]{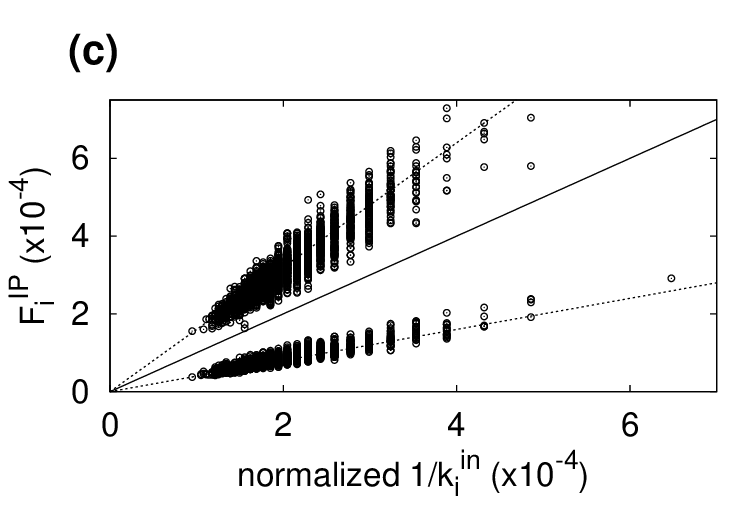}
\includegraphics[height=6cm,width=6cm]{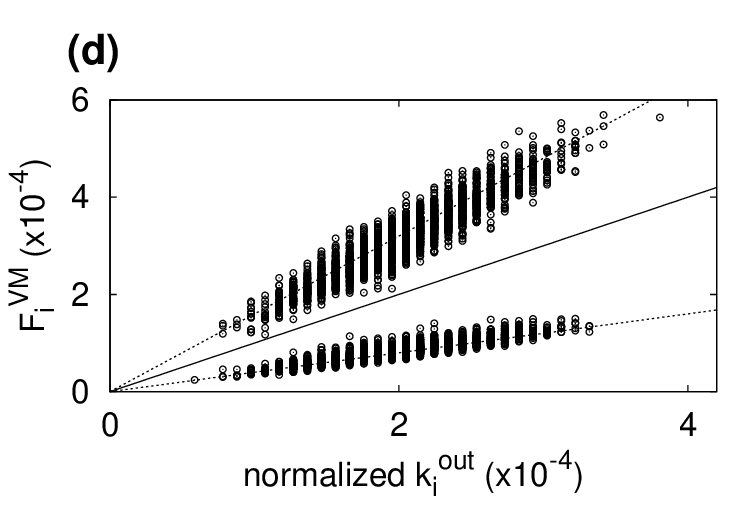}
\caption{(a) Example of modular network, with $w_{11}$,
$w_{12}$, $w_{21}$, and $w_{22}$ indicating edge weights when this
network is coarse grained as two-node network.
(b) $F_i^{LD}$, (c) $F_i^{IP}$, and
(d) $F_i^{VM}$ in an asymmetric modular network with 
$N=5000$. The solid lines represent the meanfield ansatz 
(see the captions of \FIGS\ref{fig:LD}, \ref{fig:IP}, and \ref{fig:VM}
for details). The dashed lines represent the ansatz derived from the
combination of the
local degree and the module membership of the node (see
\EQ\eqref{eq:modular ansatz} for the expression in case of LD).} 
\label{fig:init-mo}
\end{center}
\end{figure}

To intuitively understand the importance of
the global structure of networks such as community structure 
in evolutionary dynamics, we generate a modular network \cite{Leicht08prl}, as schematically shown in \FIG\ref{fig:init-mo}(a), and study the fixation probability under neutrality, \textit{i.e.}, $r=1$.
We generate two
modules $M_1=\{v_1,\ldots,v_{N/2}\}$ 
and $M_2=\{v_{N/2+1},\ldots,v_N\}$ 
as two directed random graphs with
$N/2=2500$ nodes and the mean degree $\left<k\right>_M=10$. Then, we
randomly connect $M_1$ and $M_2$ by directed edges so that a node in
$M_1$ ($M_2$) has $w_{1\to 2}\left<k\right>_M$ ($w_{2\to
1}\left<k\right>_M$) outgoing edges to the nodes in $M_2$ ($M_1$) on
an average. By setting $w_{1\to 2}=0.04$ and $w_{2\to 1}=0.01$, we obtain
a network with $N=5000$ and $\left<k\right>=10.25$. Note that the degree
correlation is absent in this network. 
For the realized network,
$F_i^{LD}$ for $r=1$ 
is shown in \FIG\ref{fig:init-mo}(b).  Rather
than the MF ansatz $\propto k_i^{out}/k_i^{in}$ (solid line),
the module membership is the main determinant of
$F_i^{LD}$. The upper and lower sets of points in
\FIG\ref{fig:init-mo}(b) correspond to the nodes in $M_1$ and $M_2$,
respectively.
The magnitude of $F_i^{LD}$ in the two sets differ
approximately by a factor of $w_{1\to 2}/w_{2\to 1}=4$.  The results
obtained in the case of the IP and VM are similar, as shown in 
\FIGS\ref{fig:init-mo}(c) and \ref{fig:init-mo}(d), respectively.
There, gross connectivity among
modules, not local degrees, principally determines $F_i$.
An modified ansatz that combines the MF approximation and the 
multiplicative factor determined by the module membership of the node
\begin{equation}
F_i^{LD}\propto \left\{\begin{array}{ll}
w_{1\to 2}k_i^{out}/k_i^{in}, & v_i\in M_1\\
w_{2\to 1}k_i^{out}/k_i^{in}, & v_i\in M_2
\end{array}\right.
\label{eq:modular ansatz}
\end{equation}
fits the data well (the dashed lines in \FIG\ref{fig:init-mo}(b)).
Similar approximations in which 
$k_i^{out}/k_i^{in}$ in
\EQ\eqref{eq:modular ansatz} is replaced with
$1/k_i^{in}$ and $k_i^{out}$ (dashed lines in 
\FIGS\ref{fig:init-mo}(b) and \ref{fig:init-mo}(c), respectively)
roughly agree with the observed $F_i^{IP}$ and $F_i^{VM}$.

To explain this result analytically, we presume that all nodes in
a module are equivalent and have an identical fixation
probability, $\hat{F}_1$ or $\hat{F}_2$.  In this manner, a network with
two modules is reduced to a network with two nodes and self-loops.
 Equation~\eqref{eq:dual LD} with $N=2$ yields
\begin{eqnarray}
\hat{F}_1^{LD} &=& \frac{w_{12}}{w_{12}+w_{21}},\\
\hat{F}_2^{LD} &=& \frac{w_{21}}{w_{12}+w_{21}}.
\end{eqnarray}
Because
$w_{11}=w_{22}=\left<k\right>_M$, $w_{12}=w_{1\to 2}\left<k\right>_M$,
and $w_{21}=w_{2\to 1}\left<k\right>_M$, we obtain
\begin{equation}
\frac{\hat{F}_1^{LD}}{\hat{F}_2^{LD}}=\frac{w_{1\to 2}}{w_{2\to 1}},
\end{equation}
which agrees with the numerical results. 
On the other hand, $k_i^{out}/k_i^{in}$ is equal to
$(1+w_{1\to 2})/(1+w_{2\to 1})$ and $(1+w_{2\to 1})/(1+w_{1\to 2})$
for $M_1$ and $M_2$, respectively.  Both of these values are close to
unity when $w_{1\to 2}, w_{2\to 1}\ll 1$, \textit{i.e.},
when the network is
modular.  Therefore, the MF approximation gives $\hat{F}_1^{LD}/
\hat{F}_2^{LD}\approx 1$, which is different from 
our simulated results. 

Similar calculations in the case of the IP
yield
\begin{equation}
\hat{F}_i^{IP}=\frac{C^{IP}_i}{C^{IP}_1+C^{IP}_2},\quad (i=1,2)
\end{equation}
where
\begin{eqnarray}
C^{IP}_1 &\equiv& \frac{w_{12}}{w_{11}+w_{12}},\\
C^{IP}_2 &\equiv& \frac{w_{21}}{w_{21}+w_{22}}.
\end{eqnarray}
Therefore, we obtain
\begin{equation}
\frac{\hat{F}_1^{IP}}{\hat{F}_2^{IP}}
\approx \frac{w_{1\to 2}}{w_{2\to 1}}
\end{equation}
when $w_{1\to 2}, w_{2\to 1}\ll 1$. However, the naive MF 
ansatz (\EQ\eqref{eq:IP MF}) yields
$1/k_i^{in}=1/\left(\left<k\right>(1+w_{2\to
1})\right)\approx 1/\left<k\right>$
($1\le i\le N/2$)  and
$1/k_i^{in}=1/\left(\left<k\right>(1+w_{1\to 2})\right)
\approx 1/\left<k\right>$ 
($(N/2)+1\le i\le N$). Then,
$\hat{F}_1^{IP}/\hat{F}_2^{IP}$ would be approximately unity,
which does not well 
explain the simulation results shown in \FIG\ref{fig:init-mo}(c).

In the case of the VM, we obtain
\begin{equation}
\hat{F}_i^{VM}=\frac{C^{VM}_i}{C^{VM}_1+C^{VM}_2},
\end{equation}
where
\begin{eqnarray}
C^{VM}_1 &\equiv& \frac{w_{12}}{w_{12}+w_{22}},\\
C^{VM}_2 &\equiv& \frac{w_{21}}{w_{11}+w_{21}}.
\end{eqnarray}
When $w_{1\to 2},w_{2\to 1}\ll 1$,
we obtain
\begin{equation}
\frac{\hat{F}_1^{VM}}{\hat{F}_2^{VM}}\approx
\frac{w_{1\to 2}}{w_{2\to 1}}.
\end{equation}
However, the naive MF ansatz (\EQ\eqref{eq:VM MF})
yields $k_i^{out}=\left<k\right>(1+w_{1\to 2})
\approx \left<k\right>$ ($1\le i\le N/2$) and
$k_i^{out}=\left<k\right>(1+w_{1\to 2})\approx
\left<k\right>$ ($N/2+1\le i\le N$). Then,
$\hat{F}_1^{VM}/\hat{F}_2^{VM}$ would be approximately unity,
which again does not
explain the simulation results shown in \FIG\ref{fig:init-mo}(d).

In sum, for each update rule the
community structure of networks
has a strong impact on the fixation probability.

\section{Conclusions}

In summary, we obtained
general formulae for the
fixation probability in directed and weighted networks.
For each of the three different update
rules, fixation probability is a solution to a system of linear
equations. 
Fixation probability in undirected networks is completely
determined by the local connectivity \cite{Antal06prl,Sood08pre}
under neutrality. In contrast,
in the case of directed
degree-correlated, small-world, or modular networks,
fixation probability is not determined only by the degree of the node
that a mutant initially invades, and it deviates
from the MF approximation to a large extent.
Our results indicate that the global connectivity of networks has a significant effect on the fixation probability.


%

\section*{Acknowledgments}

We thank Hiroshi Kori for his valuable discussions.
N.M. acknowledges the support through
the Grants-in-Aid for Scientific Research
(Nos. 20760258 and 20540382) from MEXT, Japan.
H.O. acknowledges the support through
the Grants-in-Aid for Scientific Research 
from JSPS, Japan.

\newpage
\clearpage

\clearpage

\clearpage

\clearpage

\clearpage

\clearpage

\clearpage

\end{document}